\documentclass[11pt,twoside]{article}
\usepackage{./asp2014}

\aspSuppressVolSlug
\resetcounters

\begin{document}

\title{Star Formation \& Stellar Evolution: Future Surveys \& Instrumentation}
\author{Christopher J. Evans$^1$
\affil{$^1$UK Astronomy Technology Centre, Royal Observatory, Blackford Hill, Edinburgh, EH9 3HJ, UK; 
\email{chris.evans@stfc.ac.uk}}}

\paperauthor{Christopher J. Evans}{chris.evans@stfc.ac.uk}{ }{UK Astronomy Technology Centre}{Royal Observatory Blackford Hill}{Edinburgh}{}{EH9 3HJ}{UK}

\begin{abstract}
  The next generation of multi-object spectrographs (MOS) will deliver
  comprehensive surveys of the Galaxy, Magellanic Clouds and nearby
  dwarfs. These will provide us with the vast samples, spanning the
  full extent of the Hertzsprung--Russell diagram, that are needed to
  explore the chemistry, history and dynamics of their host systems.
  Further ahead, the Extremely Large Telescopes (ELTs) will have
  sufficient sensitivity and angular resolution to extend stellar
  spectroscopy well beyond the Local Group, opening-up studies of the
  chemical evolution of galaxies across a broad range of galaxy types
  and environments. In this contribution I briefly reflect on current and
  future studies of stellar populations, and introduce plans for the
  MOSAIC instrument for the European ELT.
\end{abstract}

\section{Introduction}
Considerable breakthroughs in studies of stellar populations have been
enabled over the past 15 years by the development of high-multiplex
optical spectrographs, such as AAT-2dF (Lewis et al. 2002), VLT-FLAMES
(Pasquini et al. 2002), and Magellan-IMACS (Dressler et al. 2011).
These powerful instruments have typically been used to compile large
samples of stellar spectra to address questions in stellar evolution
(e.g., Evans et al. 2005, 2011), or to use stars as tracers of the
dynamics and assembly histories of galaxies (e.g., Zoccali et al.
2014).

This contribution highlights some of the plans for new multi-object
spectrographs (MOS) in the coming decade, and some of the
opportunities they will bring for studies of stellar populations.
Sections~\ref{IR} and \ref{next} considers recent developments for
4-10\,m class observatories, while Section~\ref{elts} looks foward to
future MOS observations with the European Extremely Large Telescope
(E-ELT), which was recently approved for construction by ESO Council
(de Zeeuw, Tamai \& Liske, 2014).

\section{Boldly into the near-IR with new technologies}\label{IR}
One of the most exciting developments in the past couple of years has
been the arrival of the first near-IR MOS instruments on large
telescopes: Keck-MOSFIRE (McLean et al. 2012) and VLT-KMOS (Sharples
et al. 2013). Although employing different technologies/approaches,
i.e., a configurable cryogenic slit unit cf.~deployable integral field
units (IFUs), both are highly capable instruments for stellar studies,
enabling efficient collection of relatively large spectroscopic
samples in the near-IR for the first time.

\subsection{Red supergiants as cosmic abundance probes}\label{rsgs}
As an example of the new research enabled by access to near-IR MOS
observations, I highlight recent studies of the physical properties
and chemical abundances of red supergiants (RSGs), the cool, luminous
descendants of massive stars. The potential of $J$-band spectroscopy
of RSGs to determine chemical abundances in galaxies was introduced by
Davies, Kudritzki \& Figer (2010) in their analysis of archival
spectra from the IRTF library (Rayner, Cushing \& Vacca, 2009). The
spectral window used in this approach is shown in Fig.~\ref{DKF10},
and has been further validated by studies of RSGs in the Galaxy (Gazak
et al. 2014) and in the Magelllanic Clouds (Davies et al. 2015).

\articlefigure{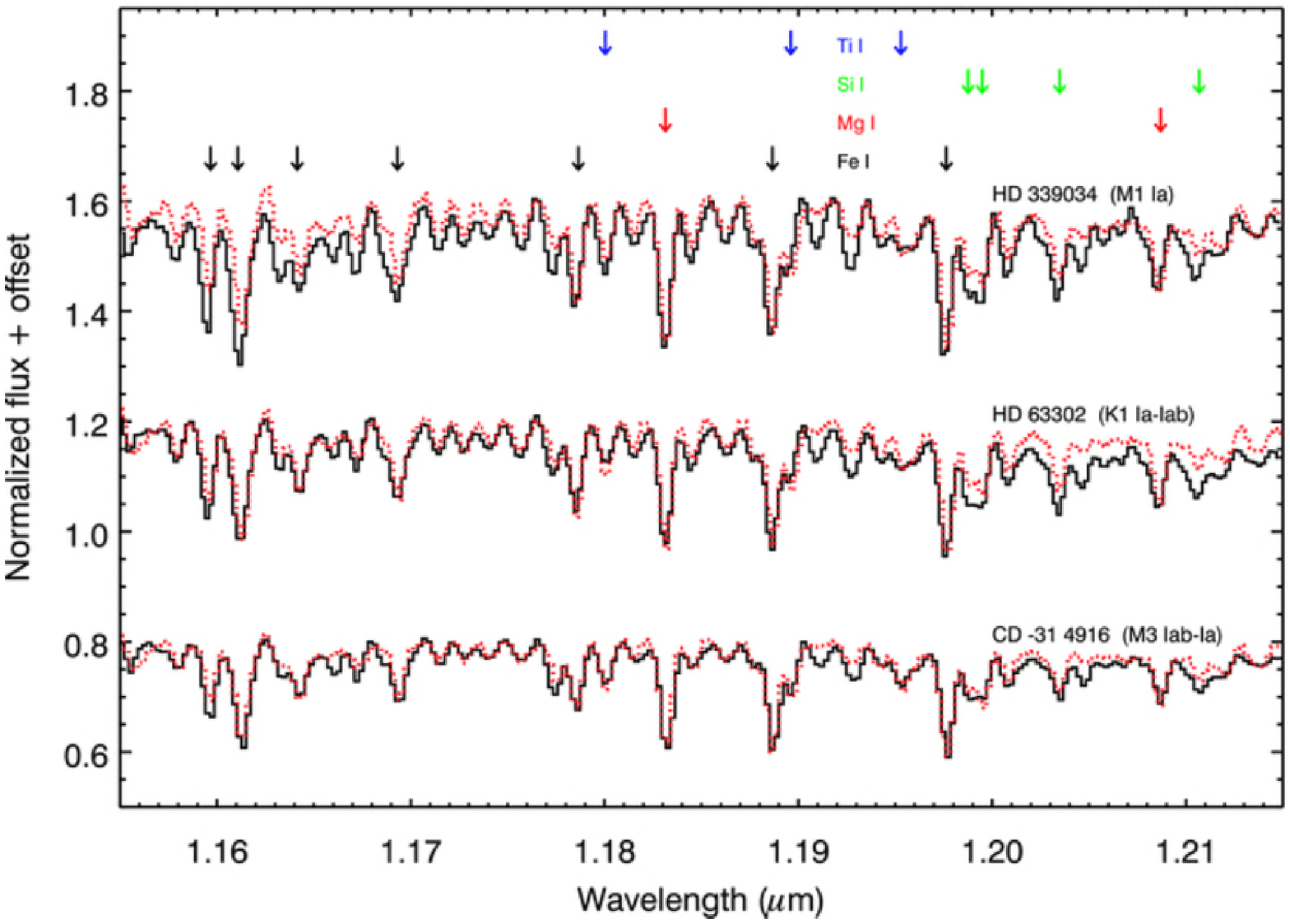}{DKF10}{Illustrative fits (dotted lines)
  to $J$-band spectra of RSGs from Davies, Kudritzki \& Figer (2010).
  This region is relatively free of telluric and sky-emission lines
  and contains useful diagnostic atomic features.}

A first application of this technique at larger distances was
demonstrated by KMOS Science Verification observations of eleven RSGs
in NGC\,6822 (d\,$=$\,0.46\,Mpc), with an estimated mean metallicity,
[$Z$], of $-$0.52\,$\pm$\,0.21 (Patrick et al. 2015). The final test
phase of this technique was KMOS Guaranteed Time Observations in
NGC\,300 (d\,$\sim$\,1.9\,Mpc), enabling a comparison of the RSG
abundances (Gazak et al. 2015, see Fig.~\ref{ZKE15}) with those for blue
supergiants from Kudritzki et al. (2008). With the method now tested
rigorously in the local Universe, efforts are underway to observe a
larger sample of galaxies (spanning a range of masses) to obtain a
direct calibration of the mass-metallicity relation (see, e.g., Kewley
\& Ellison, 2008).

In addition to observational factors such as sky and telluric
subtraction, quantitative stellar work in the near-IR also presents
new challenges in the sense that much of the focus on atomic data has
traditionally been at optical wavelengths, so new calculations are
required to, for instance, account for deviations from the
approximation of local thermodynamic equilibrium (Bergemann et al.
2012; 2013; 2015). Continued efforts will be required in this area if
we are to obtain the maximum benefit from future facilities, both from
the 8-10\,m class telescopes and the ELTs.

\articlefigure{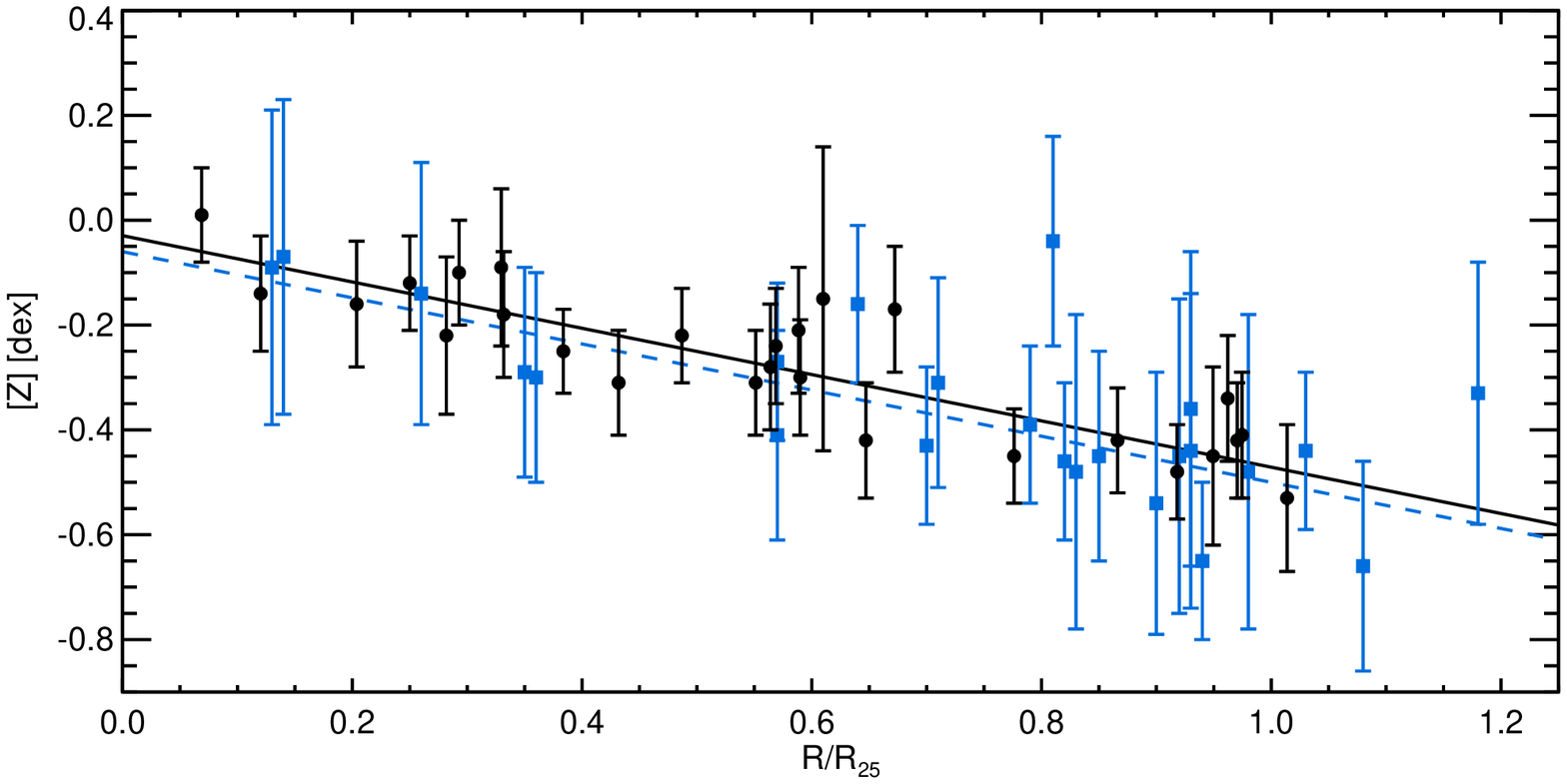}{ZKE15}{Stellar metallicities and
  gradients in NGC\,300 (Gazak et al. 2015). The results from optical
  spectroscopy of blue supergiants (Kudritzki et al. 2008, shown in
  blue) and the near-IR observations of RSGs (in black) are in
  excellent agreement.}

\section{Spectroscopy of stellar populations in 2020}\label{next}

Beyond the facilities already in operation on ground-based telescopes,
there are four MOS projects in the design/construction phase that each
have `legacy'-style large surveys of stellar populations as part of
their core programmes:
\begin{itemize}
\item{4MOST: Optical MOS for the 4\,m VISTA telescope (de Jong, this vol.);}
\item{MOONS: (red-)optical/near-IR MOS for the VLT (Cirasuolo, this vol.);}
\item{PFS: Optical/near-IR MOS for the Subaru Telescope (Takada, this vol.);}
\item{WEAVE: Optical MOS for the WHT (Dalton, this vol.).}
\end{itemize}

The surveys enabled by this next generation of instruments will
provide a truly vast ($>>$10$^6$ stars) census of the stellar
populations of the inner Milky Way, its disk and halo populations, the
Magellanic Clouds, and the dwarf/irregular galaxies of the Local
Group. Other instruments under construction such as GTC-MEGARA (de
Paz, this vol.) and GTC-EMIR (Garzon, this vol.) will also contribute
valuable observations.

Many of the detailed plans for the stellar surveys with these new
instruments are presented elsewhere in this volume. As an example of
studies of star formation (specifically whether it always occurs in
dense clusters), spatial analyses of star-forming regions can already
provide us with insights into their initial conditions and dynamical
evolution (e.g., Parker et al. 2014; Wright et al. 2014), but the
information encoded in the radial velocities (from the spectroscopy)
and proper motions (from the {\it Gaia} mission) will give us unique
three-dimensional information to trace their dynamical histories (e.g.,
to identify comoving groups with the different spatial
sub-structures).

\section{The Science Case for an ELT-MOS}\label{elts}

\begin{center}
\indent{\it `There's a capacity for appetite
      that a whole heaven and earth of cake can't satisfy'\\John
      Steinbeck (East of Eden)}
\end{center}

The MOS facilities planned for the coming years will be transformative
due to the combination of their large multiplexes with substantial
telescope allocations to acquire the vast samples required. However,
as we look further ahead, we are ultimately limited by the collecting
area of current facilities. For instance, Keck-DEIMOS spectroscopy of
the evolved stellar populations in M31 gives insufficient
signal-to-noise below the tip of the red giant branch (at
$I$\,$>$\,21.5\,mag, e.g., Chapman et al. 2006), and quantitative
analysis of massive O-type stars is limited to all but the most
luminous objects beyond 1\,Mpc (e.g., Tramper et al. 2011, 2014).

The ELTs will provide a huge leap forward in both sensitivity and, via
adaptive optics (AO), spatial resolution. In addition to ESO's planned
39\,m E-ELT, there are two other ELT projects now entering the
construction phase: the Giant Magellan Telescope (GMT) and the Thirty
Meter Telescope (TMT). Some of the challenges of simply scaling-up
current MOS designs and capabilities to ELT-class instruments are
discussed by Bernstein (this vol.). Nonetheless, there is a huge range
of scientific topics which require ELT-MOS observations, ranging from
spectroscopic characterisation of the most distant galaxies, through
to studies of exoplanets in stellar clusters (Evans et al. 2015).

\subsection{MOSAIC: The MOS for the E-ELT}
Following Phase~A studies of three potential E-ELT MOS concepts (see
Ramsay et al. 2010), European and Brazilian astronomers have combined
their efforts to assemble a comprehensive science case for an ELT-MOS
(Evans et al. 2015), working together on the MOSAIC instrument concept
(Hammer et al. 2014).

The range of cases presented by Evans et al. (2015) flow down to
instrument requirements which are knowingly broad (see their Table~7).
This step was intended as a first census of all the potential cases
and relevant parameter space for MOS observations. The Phase~A
conceptual design of MOSAIC is anticipated to start in late 2015,
including scientific trade-offs of capability vs.~cost (and technical
feasibility). One of the key approaches in plans for MOSAIC has been
the delineation of possible MOS sources into two types of observations,
identified by Evans et al. (2012) as:
\begin{itemize}
\item{{\it High definition:} Observations of tens of channels at fine
    spatial resolution, with multi-object adaptive optics (MOAO) providing
    high-performance correction for selected sub-fields.}
\item{{\it High multiplex:} Integrated-light (or coarsely-resolved)
    observations of $>$100 objects at the spatial resolution given 
    by the ground-layer adaptive optics (GLAO) of the telescope.}
\end{itemize}

The technical readiness of the high-definition mode has advanced
significantly over the past few years thanks to the CANARY project on
the WHT in La Palma. This has performed the first on-sky
demonstrations of MOAO using natural guide stars (Gendron et al. 2011;
Vidal et al. 2014) and, more recently, using laser guide stars. As
illustrated by the point-spread functions in Fig.~\ref{VGR14}, while
MOAO does not yield the same performance as single-conjugate AO, it is
substantially better than that from seeing or GLAO. The attraction of
this approach is that such performance can potentially be obtained for
multiple sub-fields within the large ($\sim$10$'$ diameter) field of
the E-ELT.

A second MOAO pathfinder, RAVEN on the Subaru Telescope, has also made
impressive progress with on-sky tests in the past year (Lardi\`{e}re
et al. 2014). As an aside, note that even in the deep cosmological
fields that are deliberately free of bright foreground stars, there
are still sufficient (fainter) stars for significant improvements in
image quality from MOAO (Basden, Evans \& Morris, 2014).

\articlefigure{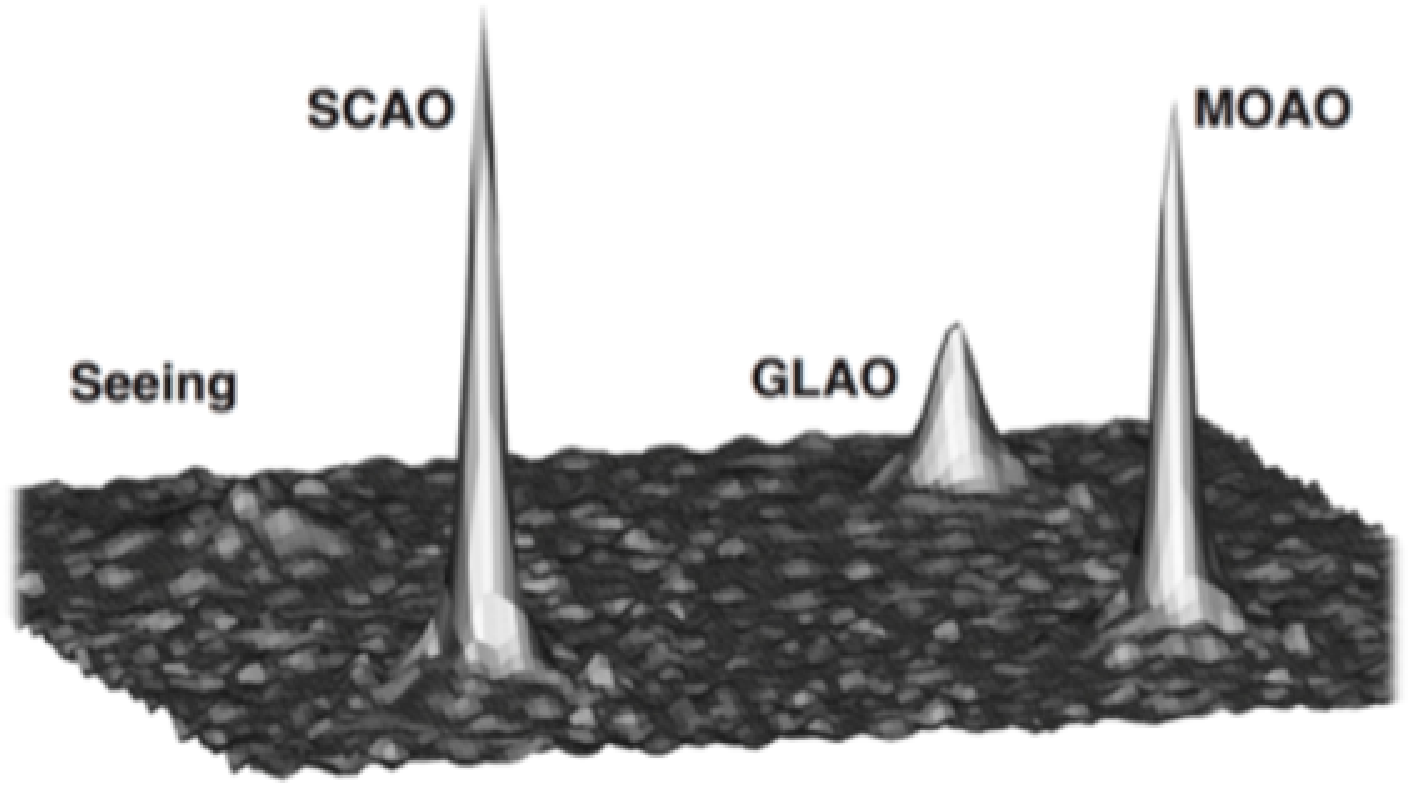}{VGR14}{Example $H$-band point-spread
  functions from the CANARY experiment (Vidal et al. 2014), obtained
  without AO correction (seeing) and with ground-layer, multi-object,
  and single-conjugate AO (GLAO, MOAO, and SCAO, respectively).}

\subsection{Resolved stellar populations beyond the Local Group}

A key component of the MOSAIC case, indeed for the ELTs in general, is
concerned with stellar populations, both in the inner Milky Way and in
distant systems beyond 1\,Mpc. For example, when combined with AO, the
$J$-band method introduced in Section~\ref{rsgs} is potentially very
powerful, opening-up direct abundance studies of RSGs out to distances
of tens of Mpc (Evans et al. 2011). Work is now underway to explore
this technique for lower-mass, evolved stars on the red giant branch,
to compare it with the use of the Calcium triplet in estimation of
stellar metallicities and radial velocities (e.g. Tolstoy et al.
2001; Battaglia et al. 2008).

An expanded case for this topic was presented by Evans et al. (2015).
In short, MOSAIC spectroscopy will open-up studies of the evolved
populations in galaxies such as the spirals in the Sculptor Group for
the first time. If employing {\it high definition} AO-corrected
observations with IFUs, note that the effective multiplex of stars
observed in dense regions is likely to be much larger than simply the
number of IFUs, potentially giving samples of 1000s of stars within a
relatively modest amount of time. Equally, {\it high multiplex}
optical spectroscopy would provide high-quality observations of
massive O-type stars to investigate their physical properties in
galaxies beyond 1\,Mpc and/or studies of the evolved stars in the
interesting halo regions of these distant systems.

\acknowledgements My thanks to the organisers for their kind
invitation to speak at the conference, to Ben Davies and Nick Wright
for helpful chats ahead of the meeting, and to the staff of La
Cuatro for a friendly welcome even after the best part of a decade.

\end{document}